\documentclass[12pt]{article}
\usepackage{graphicx}
\usepackage{epsfig}

\begin{document}
\topskip 2 cm 
\renewcommand{\thefootnote}{\fnsymbol{footnote}}  

\begin{titlepage}
\rightline{ \large{ \bf April 2002} }
\begin{center}
{\Large\bf The analytical values of the sunrise master integrals  }\\ 
{\Large\bf for one of the masses equal to zero. }\\

\vspace{2.5cm}
{\large {\bf
H.~Czy{\.z}\ , }}
{\large {\bf
A.~Grzeli{\'n}ska\ }}
{\large and   
{\bf R.~Zabawa  \\ } }

%\vspace{.5cm}
\begin{itemize}
\item[]{\sl Institute of Physics, University of Silesia, 
             PL-40007 Katowice, Poland }
\end{itemize}
\end{center}

\noindent
e-mail: {\tt  czyz@us.edu.pl \\ 
\hspace*{1.2cm} grzel@joy.phys.us.edu.pl \\ 
 \ } 
\vspace{.5cm}
% \vspace{2.5cm}
% \vfil
\begin{center}
\begin{abstract}

 The analytical values of the sunrise master amplitudes were found
 for one mass equal to zero, two other arbitrary masses and an arbitrary
 external four momentum. Differential equations
 in the square of the external four momentum and masses were used to
 obtain the master integrals. 

\end{abstract}
\end{center}

\vskip  1cm

{\scriptsize \noindent ------------------------------- \\ }
\noindent
{\it keywords:} sunrise diagram, analytical calculations, master integrals.
% \small{ \noindent 

\scriptsize{ 
} 
\vfill
\end{titlepage}
\pagestyle{plain} \pagenumbering{arabic}

%%%%%%%%%%%%%%%%%%%%%%%%%%%%%%%%%%%%%%%%%%%%%%%%%%%%%%%%%%%%%%%%%%%%%%%% 
\newcommand{\Eq}[1]{Eq.(\ref{#1})} 
\newcommand{\labbel}[1]{\label{#1}} 
\newcommand{\cita}[1]{\cite{#1}} 
\def\Re{\hbox{Re~}} 
\def\Im{\hbox{Im~}} 
\newcommand{\F}[1]{F_#1(n,m_1^2,m_2^2,m_3^2,p^2)} 
\newcommand{\dnk}[1]{ d^nk_{#1} } 
\newcommand{\Fn}[2]{F_{#1}^{(#2)}(m_1^2,m_2^2,m_3^2,p^2)}
\newcommand{\D}{D(m_1^2,m_2^2,m_3^2,p^2)} 
\def\Li2{\hbox{Li}_2}
%\def\a{\alpha} 
%\def\app{{\left(\frac{\alpha}{\pi}\right)}} 
%\newcommand{\e}{{\mathrm{e}}} 
%%%%%%%%%%%%%%%%%%%%%%%%%%%%%%%%%%%%%%%%%%%%%%%%%%%%%%%%%%%%%%%%%%%%%%%% 

\section{Introduction.} 

 High precision measurements in elementary particle physics require
 more and more precise calculations of multi-loop Feynman diagrams
 allowing in the same time detailed tests of the existing models.

 The widely used integration by part technique \cite{TkaChet} allows for
 reduction of the complicated task of calculation of radiative corrections 
 to calculation of a limited number of scalar integrals called master
 integrals (MI). Whenever possible it is of great importance to know
 analytically that basic building blocks, as it safes enormously 
 the CPU time required for their numerical evaluation. 
 Investigations of the sunrise two-loop diagram started a long time ago
 and it is beyond the scope of that paper to present the complete
 literature of the subject, so we will give here only a partial
 guideline of the literature sending the reader to the references
 of the quoted paper for further studies.

 The MI of the sunrise graph were found to be expressible as a 
 combination of Lauricella functions in \cite{BBBS}. Their analytical
 expansions at \(p^2=0\)  and \(p^2=\infty\) were given in
 \cite{BBBS} and \cite{CCLR1}. Their values at threshold 
 \(p^2 = -(m_1+m_2+m_3)^2\) and pseudothresholds (\(p^2 = -(m_1-m_2+m_3)^2\),
\(p^2 = -(m_1+m_2-m_3)^2\), \(p^2 = -(-m_1+m_2+m_3)^2\) ) were 
 presented in \cite{BDU}.
 The analytical expansions at pseudothresholds were found in \cite{CCR1}.
 A semi-analytical expansion at threshold was given 
 in \cite{DS} and also using the
 configuration space technique in \cite{GP}, while the complete analytical
 expansion at threshold was presented in \cite{CCR2}. Some special
 values of the sunrise MI were obtained in \cite{FKK} and \cite{AMR}.

%%%%%%%%%%%%%%%%%%%%%%%%%%%%%%%%%%%%%%%%%%%%%%%%%%%%%%%%%%%%%%%%%%%%%%%%%
\section{Calculation of the sunrise master integrals.} 

The scalar integrals related to the sunrise two-loop graph can be defined as
\cite{CCLR1}

\begin{eqnarray} 
  A(n,m_1^2,m_2^2,m_3^2,p^2,-\alpha_1,-\alpha_2,-\alpha_3,\beta_1,\beta_2) 
      &=& \frac{ 1}{((2\pi)^{n-2})^2 } \nonumber \\  
      && {\kern-220pt} \int \dnk{1} \int \dnk{2} \; 
      \frac{ (p\cdot k_1)^{\beta_1} (p\cdot k_2)^{\beta_2} } 
           { (k_1^2+m_1^2)^{\alpha_1} (k_2^2+m_2^2)^{\alpha_2} 
             ( (p-k_1-k_2)^2+m^2_3 )^{\alpha_3} } \ \ ,
 \nonumber \\
\labbel{1} \end{eqnarray} 

\noindent
where \(m_i\) (\(i=1,2,3\) ) are the masses associated with internal lines,
\(p\) is the external momentum, \(k_1,k_2\) are loop momenta and 
 \(\alpha_i,\ \ (i=1,2,3)\), \(\beta_j,\ \ (j=1,2)\) are integer numbers.
 The integrals are to be performed in \(n\)-dimensional Euclidean space.
 It means we use the dimensional regularization and have performed 
 the Wick rotation.
 The scale parameter \(\mu\) associated with the dimensional regularization
 was set to 1. Final results can be easily rewritten in Minkowski space
 by changing \(p^2\rightarrow-p^2\).

 The four independent Master Integrals (MI) were chosen as 

\begin{eqnarray} 
 \F{j}
      &=& \frac{ 1}{((2\pi)^{n-2})^2 }\nonumber \\  
      && {\kern-120pt} \int \dnk{1} \int \dnk{2} \; 
      \frac{ 1 } 
           { (k_1^2+m_1^2)^{\alpha_1(j)} (k_2^2+m_2^2)^{\alpha_2(j)} 
             ( (p-k_1-k_2)^2+m_3^2 )^{\alpha_3(j)} } \ , 
\labbel{2} \end{eqnarray} 

\noindent
with $ j=0,1,2,3;$ $ i=1,2,3;$ $\alpha_i(0)=1$; 
 \(\alpha_i(j)=1\), for \(j\ne i\);
 \(\alpha_i(j)=2\), for \(j = i\). While their \((n-4)\) expansions 
 are defined \cite{CCLR1} by

\begin{eqnarray} 
F_j(n,m_1^2,m_2^2,m_3^2,p^2) &=&
 \frac{1}{(n-4)^2}F_j^{(-2)}(m_1^2,m_2^2,m_3^2,p^2)
+\frac{1}{(n-4)}F_j^{(-1)}(m_1^2,m_2^2,m_3^2,p^2)\nonumber \\ 
&+& F_j^{(0)}(m_1^2,m_2^2,m_3^2,p^2) + \cdots \ , \ \ \  \  j=0,1,2,3
 \labbel{expm} 
\end{eqnarray}

The calculation of the MI for \(m_3=0\) presented here
 make use of two systems of differential equations
satisfied by the MI of the sunrise diagram
given in \cite{CCLR1}. One of them in the variable \(p^2\)

\begin{eqnarray} 
p^2 {\frac{\partial} {\partial p^2}} F_0 &=&
(n-3)F_0 + m_1^2F_1+m_2^2F_2+m_3^2F_3 \nonumber \\ 
 p^2 \ D(m_1^2,m_2^2,m_3^2,p^2) 
  \ {\frac{\partial}{\partial p^2}} F_i &=&
 \sum_{j=0}^{3} M_{i,j}  F_j + T_i  \ \ , \ i=1,2,3 \ \ , 
 \labbel{a5}
 \end{eqnarray}

\noindent
 where the explicit form of functions \(T_i\) (expressed as functions of
 \(T\) ) 
 and \(M_{i,j}\) (polynomials of 
\(p^2, m_1^2,m_2^2,m_3^2\) ) can be found in \cite{CCLR1}.
 The function \(D\) is defined by

 \begin{eqnarray} 
D(m_1^2,m_2^2,m_3^2,p^2)=
  && \left(p^2+(m_1+m_2+m_3)^2\right) \ 
                        \left(p^2+(m_1+m_2-m_3)^2\right) \nonumber \\ 
     &&\kern-6pt \cdot \left(p^2+(m_1-m_2+m_3)^2\right) \ 
                        \left(p^2+(m_1-m_2-m_3)^2\right) \ ,
  \labbel{a7}
 \end{eqnarray} 

 and the aforementioned function \(T\)  by 

\begin{eqnarray} 
 T\left(n,m^2\right)=
 \int \frac{d^n k}{(2 \pi)^{n-2}} \frac{1}{k^2+m^2}
 = {\frac{m^{n-2}}{(n-2)(n-4)}} C(n) \ \ ,
 \labbel{a1}
 \end{eqnarray} 

\noindent
where

\begin{eqnarray} 
 C(n)= \left(2 \sqrt{\pi} \right)^{(4-n)} \Gamma\left(3-\frac{n}{2}\right) 
 \ \ \ \  {\mbox {\rm and}} \ \ \ \   C(4)=1 \ \ .
\labbel{a11}
 \end{eqnarray}

The second system of differential equations, with \(m_3^2\) as a variable,
can be easily obtained using expressions
 presented in the Appendix of \cite{CCLR1}. They represent

\noindent
 \(  A(n,m_1^2,m_2^2,m_3^2,p^2,-3,-1,-1,0,0) \) 
and \( A(n,m_1^2,m_2^2,m_3^2,p^2,-1,-2,-2,0,0)\)
as linear 

\noindent
combinations of the MI. Remembering that 

\begin{eqnarray}
A(n,m_1^2,m_2^2,m_3^2,p^2,-1,-2,-2,0,0)&=& 
-\frac{\partial F_{2}(n,m_{1}^2,m_{2}^2,m_{3}^2,p^2) }{\partial m_3^2}
 \nonumber \\
A(n,m_1^2,m_2^2,m_3^2,p^2,-1,-1,-3,0,0) &=& -\frac{1}{2}
\frac{\partial F_{3}(n,m_{1}^2,m_{2}^2,m_{3}^2,p^2) }{\partial m_3^2}
\nonumber \\
etc.
\labbel{ee1} 
\end{eqnarray}

\noindent
 they are also differential equations in the variable \(m_3^2\)
 (or in any other mass by means of permutation of masses).
 They can be written as

\begin{eqnarray}
\frac{\partial}
 {\partial m_{3}^2}F_{0}(n,m_{1}^2,m_{2}^2,m_{3}^2,p^2) 
= A_{0,0}F_{0}(n,m_{1}^2,m_{2}^2,m_{3}^2,p^2)
 + A_{0,1}F_{1}(n,m_{1}^2,m_{2}^2,m_{3}^2,p^2) \nonumber \\
+A_{0,2}F_{2}(n,m_{1}^2,m_{2}^2,m_{3}^2,p^2)
+ A_{0,3}F_{3}(n,m_{1}^2,m_{2}^2,m_{3}^2,p^2)+ B_{0}  \nonumber\\
\nonumber \\
\frac{\partial}{\partial m_{3}^2}F_{i}(n,m_{1}^2,m_{2}^2,m_{3}^2,p^2) = 
A_{i,0}F_{0}(n,m_{1}^2,m_{2}^2,m_{3}^2,p^2) 
+ A_{i,1}F_{1}(n,m_{1}^2,m_{2}^2,m_{3}^2,p^2)~~~\nonumber \\
+A_{i,2}F_{2}(n,m_{1}^2,m_{2}^2,m_{3}^2,p^2)
+ A_{i,3}F_{3}(n,m_{1}^2,m_{2}^2,m_{3}^2,p^2)+ B_{i} \ \ ,  \nonumber\\
\labbel{sys2}
\end{eqnarray} \nonumber

\noindent
where \(i=1,2,3 \ .\) 
We do not present here for short the explicit form of the coefficients
 \(A_{i,j}\)  and \( B_{i}\) as they can be easily
found from \cite{CCLR1}.

\noindent
 To obtain the values of the MI at \(m_3^2=0\) we write the expansion of
 \(F_0(n,m_{1}^2,m_{2}^2,m_{3}^2,p^2)\) at that point 
(the expansions of the other
 master integrals are obtained using \Eq{ee1})

\begin{eqnarray}
F_0(n,m_{1}^2,m_{2}^2,m_{3}^2,p^2)
 = \sum_{\alpha_i \in A } 
 (m_{3}^2)^{\alpha_i}\cdot \sum_{k=0}^{\infty}F^{\alpha_i}_k(m_{3}^2)^k
\ \ ,
\labbel{exp1}
\end{eqnarray}

\noindent
where \(A\) is a set of all allowed values of \(\alpha_i\).
 The allowed values of \(\alpha_i\) one can easily get
putting \Eq{exp1} and its \(m_i^2\) derivatives into \Eq{sys2}.
 We have found that there exist
only two allowed values of the powers \(\alpha_i\) and the expansion of
\(F_0(n,m_{1}^2,m_{2}^2,m_{3}^2,p^2)\)  reads

\begin{eqnarray}
F_0(n,m_{1}^2,m_{2}^2,m_{3}^2,p^2)
 = \sum_{k=0}^{\infty}F^{r}_k(m_{3}^2)^k
+(m_{3}^2)^{(n-2)/2}\cdot \sum_{k=0}^{\infty}F^{s}_k(m_{3}^2)^k \ \  .
\labbel{exp2} 
\end{eqnarray}

It consists of two independent series, the regular one denoted 
 as \({r}\) and the
 singular one denoted as \(s\). It means that the 
\(m_3^2 =0 \) point is a regular singular point
 \cite{Ince} of that equations. The equations \Eq{sys2} provide also
 with the following
 differential equations 

\begin{eqnarray}
R^2(m_{1}^2,m_{2}^2,-p^2)\frac{\partial}{\partial m_{2}^2} F_{0}^{s} &=& 
(n-3)(m_{2}^2-m_{1}^2+p^2)F_{0}^{s}-\frac{C^2(n)(m_{1}^2)^{(n-2)/2}}
 {(n-2)(n-4)^2}\nonumber\\
&+&\frac{C^2(n)(m_{2}^2)^{(n-4)/2}(m_{1}^2+m_{2}^2+p^2)}{2(n-2)(n-4)^2} 
  \nonumber\\
\nonumber\\
R^2(m_{1}^2,m_{2}^2,-p^2)\frac{\partial}{\partial m_{1}^2} F_{0}^{s} &=& 
(n-3)(m_{1}^2-m_{2}^2+p^2)F_{0}^{s}
-\frac{C^2(n)(m_{2}^2)^{(n-2)/2}}{(n-2)(n-4)^2}\nonumber\\
&+&\frac{C^2(n)(m_{1}^2)^{(n-4)/2}(m_{1}^2+m_{2}^2+p^2)}
{2(n-2)(n-4)^2} \ \ , 
 \labbel{eqm1m2}
\end{eqnarray}

\noindent
where

\begin{eqnarray}
R^2(-p^2,m_{1}^2,m_{2}^2) = p^4+m_{1}^4
+m_{2}^4+2m_{1}^2p^2+2m_{2}^2p^2-2m_{1}^2m_{2}^2  \ .
\end{eqnarray}

 On the first sight it is not necessary to know the singular part when looking
for the value of the MI at \(m_3=0\) as it vanishes in that limit. 
However as it
will be clear from the discussion below knowledge of its poles in \((n-4)\) 
provides an essential simplification in the calculation of the regular part.

The equations \Eq{eqm1m2} can in principle be solved, but it is 
 much easier to use 
 system of equations \Eq{a5}  with \(p^2\) as an independent variable
 for calculation of \(F_{0}^{s}\).
 Substituting 
\(F_0(n,m_{1}^2,m_{2}^2,m_{3}^2,p^2)\)  and other MI with their mass expansions
in the system of equations \Eq{a5}
  we have found that \(F_{0}^{s} \) satisfies the following
differential equation

\begin{eqnarray}
 \kern-3pt \frac{\partial}{\partial p^2} F_{0}^{s} &=&
\frac{1}{2p^2}(n-4)F_{0}^{s}
 -\frac{1}{p^2R^2(m_{1}^2,m_{2}^2,-p^2)}
 \biggl[\hspace{1ex} (n-3)\left(p^2(m_{1}^2+m_{2}^2)
 +(m_1^2-m_2^2)^2\right)F_{0}^{s}\nonumber\\
  &+& \frac{C^2(n)}{2(n-2)(n-4)^2}
(m_{1}^2)^{(n-2)/2}(p^2-m_{2}^2+m_{1}^2) \nonumber\\
 &+& \frac{C^2(n)}{2(n-2)(n-4)^2}(m_{2}^2)^{(n-2)/2}
(p^2-m_{1}^2+m_{2}^2) \hspace{1ex}\biggr] \ .
 \labbel{eqp2}
\end{eqnarray}

We have found 
a solution of that differential equation in a form expanded around \(n=4\).
 Only pole parts of the function were calculated
  as only the poles are necessary for our purpose of calculating the
 values of the MI at \(m_3=0\).
 The solution reads

\begin{eqnarray}
&&{\kern-30pt} F_{0}^{s} = -\frac{C^2(n)}{4(n-4)^2} +
\frac{C^2(n)}{16(n-4)}\biggl[ 6 - \log(m_{1}^2)- \log(m_{2}^2)
+ \frac{m_{1}^2-m_{2}^2}{p^2}\log\left(\frac{m_{1}^2}{m_{2}^2}\right) 
 \biggr] \nonumber \\
&&+ \frac{1}{(n-4)}R(m_{1}^2,m_{2}^2,-p^2)   
 \biggl[ \frac{C^2(n)}{8p^2}\log(t) - \frac{S_{l}}{p^2} \biggr]
 + \ {\cal O}\left((n-4)^0\right)
 \hspace{3ex} \ ,  
\labbel{sol1}
\end{eqnarray}

\noindent
where \(S_l\) is an integration constant. 
The unknown function of \( m_1\) and \(m_2\), which remains after \(p^2\)
 integration,
was reduced to a single numerical constant \(S_l\) using 
the equations \Eq{eqm1m2}.

The system of equations \Eq{sys2} gives no information about
 \(F_0^r\),  provides however an expression for
the second coefficient in the regular series. It can be 
expressed as a linear combination of \(F_0^r\) and its derivatives

\begin{eqnarray}
&&{\kern-30pt} F_{1}^{r} =\frac{1}{R^2(m_{1}^2,m_{2}^2,-p^2)} 
  \Biggl(\hspace{2ex} \frac{3}{(n-4)}
(n-\frac{8}{3})(n-3)(m_{2}^2+m_{1}^2+p^2)F_{0}^{r} \nonumber \\ &&-
\frac{4}{(n-4)}m_{1}^2(n-3)(p^2+m_{1}^2)
\frac{\partial}{\partial m_{1}^2}F_{0}^{r}
 - \frac{4}{(n-4)}m_{2}^2(n-3)(p^2+m_{2}^2)
\frac{\partial}{\partial m_{2}^2}F_{0}^{r}  \nonumber \\ &&+
 \frac{1}{(n-4)^3}(m_{1}^2)^{(n-2)/2}
(m_{2}^2)^{(n-2)/2}C^2(n) \hspace{2ex}\Biggr) \ .
\labbel{F1}
\end{eqnarray}

Putting the expansion \Eq{exp2} into the equations \Eq{a5}
one finds a system of three differential equations
satisfied by \(F_{0}^{r}\), 
 \(\frac{\partial}{\partial m_{1}^2} F_{0}^{r}\) 
 and \(\frac{\partial}{\partial m_{2}^2} F_{0}^{r}\)

\begin{eqnarray} 
\frac{\partial}{\partial p^2}F_{0}^{r} &=&      
\frac{1}{p^2}(n-4)F_{0}^{r}+\frac{1}{p^2}F_{0}^{r}-
\frac{m_{1}^2}{p^2}\frac{\partial}{\partial m_{1}^2} F_{0}^{r}
-\frac{m_{2}^2}{p^2}\frac{\partial}{\partial m_{2}^2}F_{0}^{r}
  \nonumber\\
\frac{\partial}{\partial p^2} 
\frac{\partial}{\partial m_{1}^2} F_{0}^{r} &=& 
\frac{1}{R^2(m_{1}^2,m_{2}^2,-p^2)} 
 \Biggl[ \hspace{1ex} \frac{p^2 +m_{1}^2-m_{2}^2}{2p^2}
\biggl(4+7(n-4)+
3(n-4)^2 \biggr)F_{0}^{r} \nonumber \\
&+&
(n-4)\frac{\partial }{\partial m_{1}^2}F_{0}^{r}
\biggl( \frac{2m_{1}^2m_{2}^2}{p^2}-\frac{3m_{1}^4}{2p^2} 
-\frac{m_{2}^4}{2p^2}+ \frac{p^2}{2} - m_{1}^2\biggr)\nonumber\\
&+&\frac{\partial}{\partial m_{1}^2}F_{0}^{r} 
\biggl(  \frac{3m_{1}^2m_{2}^2}{p^2}
 -\frac{2m_{1}^4}{p^2} 
-\frac{m_{2}^4}{p^2} -2m_{1}^2 - m_{2}^2 \biggr)
\nonumber\\
&+&(n-3)\frac{\partial }{\partial m_{2}^2}F_{0}^{r}
\biggl(-\frac{m_{1}^2m_{2}^2}{p^2} + \frac{m_{2}^4}{p^2} 
- 3m_{2}^2\biggr)\nonumber\\
&+&\frac{C^2(n)(m_{1}^2)^{(n-4)/2}(m_{2}^2)^{(n-2)/2}
(-p^2+m_{1}^2-m_{2}^2)}{4(n-4)^2p^2} \hspace{1ex} \Biggr] \nonumber \\
\frac{\partial}{\partial p^2} 
\frac{\partial }{\partial m_{2}^2}F_{0}^{r} &=& 
\frac{1}{R^2(m_{1}^2,m_{2}^2,-p^2)}
\Biggl[ \hspace{1ex}\frac{p^2-m_{1}^2+m_{2}^2}{2p^2}
\biggl(4+7(n-4)+3(n-4)^2\biggr) F_{0}^{r}
\nonumber \\
&+&(n-4)\frac{\partial }{\partial m_{2}^2}F_{0}^{r}
 \biggl( \frac{2m_{1}^2m_{2}^2}{p^2} - \frac{m_{1}^4}{2p^2}  
- \frac{3m_{2}^4}{2p^2} +\frac{p^2}{2}- m_{2}^2 \biggr) \nonumber \\
&+&\frac{\partial }{\partial m_{2}^2}F_{0}^{r}
 \biggl( \frac{3m_{1}^2m_{2}^2}{2p^2} -\frac{m_{1}^4}{p^2} 
-\frac{2m_{2}^4}{p^2} - m_{1}^2 
- 2m_{2}^2 \biggr) \nonumber\\
&+&(n-3)\frac{\partial}{\partial m_{1}^2}F_{0}^{r}
 \biggl( -\frac{m_{1}^2m_{2}^2}{p^2} 
+\frac{m_{1}^4}{p^2} - 3m_{1}^2 \biggr) \nonumber \\
&+&\frac{C^2(n)(m_{1}^2)^{(n-2)/2}(m_{2}^2)^{(n-4)/2}
 (-p^2-m_{1}^2+m_{2}^2)}
{4(n-4)^2p^2}\hspace{1ex}  \Biggr] \ \ ,  \nonumber\\
\labbel{sys4}
\end{eqnarray}

 Trying to find \(F_0^r\) from \Eq{sys4} only
 would mean to solve a third order differential equation satisfied by 
 \(F_0^r\). 
 That task can be simplified enormously by careful
 examination of the  
\Eq{F1}.
 One finds from it that in the \((n-4)\) expansion of \(F_1^r\) the terms
 \(\sim (n-4)^0 \) of \(F_0^r\) contribute to pole terms of \(F_1^r\).
 As a result, knowing the pole terms of the singular part of 
the expansion (\Eq{sol1}) and the exact pole terms of the complete 
master integral \(F_0 \) \cite{CCLR1},
 we can find an additional relation between \(F_0^r\)
and its mass derivatives. Defining the \((n-4)\) expansion of the
 \(F_0^r\) by

\begin{eqnarray}
F_{0}^{r}=\frac{1}{(n-4)^2 }F_{0,-2}^{r}+\frac{1}{(n-4)}F_{0,-1}^{r}+F_{0,0}^{r}+{\cal O}(n-4) \ ,
\end{eqnarray}

\noindent
one finds

\begin{eqnarray} 
0=(m_{1}^2+m_{2}^2+p^2)F_{0,0}^{r}
-m_{1}^2(p^2+m_{1}^2)\frac{\partial}{\partial m_{1}^2}F_{0,0}^{r}
-m_{2}^2(p^2+m_{2}^2)\frac{\partial}{\partial m_{2}^2}F_{0,0}^{r} 
\nonumber \\
+R(m_{1}^2,m_{2}^2,-p^2)\biggl(\frac{C^2(n)}{32}\log(t)-\frac{S_{l}}{4}\biggr)
\biggl(p^2+2m_{1}^2+2m_{2}^2+\frac{1}{p^2}(m_{1}^2-m_{2}^2)^2\biggr)
\nonumber\\
+C^2(n)\Biggl( \hspace{1ex}\frac{(p^2)^2}{128}
\biggl(13-2\log(m_{1}^2)-2\log(m_{2}^2)
\biggr)+\frac{p^2}{128}\biggl(41(m_{1}^2+m_{2}^2) \nonumber \\
-\log(m_{1}^2)(14m_{1}^2+ 6m_{2}^2)-\log(m_{2}^2)(6m_{1}^2+14m_{2}^2)\biggr)
+\frac{1}{64p^2}\biggl(m_{2}^4(3m_{1}^2-m_{2}^2)\nonumber\\
-m_{1}^4(3m_{2}^2-m_{1}^2)\biggr)\biggl(\log(m_{1}^2)-\log(m_{2}^2)\biggr)
+\frac{m_{1}^2m_{2}^2}{32}\biggl(\log(m_{1}^2)
+\log(m_{2}^2)\biggr)^2\nonumber \\
+\frac{1}{32}(12m_{1}^2m_{2}^2+7m_{1}^4+7m_{2}^4)
-\frac{\log(m_{1}^2)}{64}(12m_{1}^2m_{2}^2+5m_{1}^4+3m_{2}^4)
\nonumber\\
-\frac{\log(m_{2}^2)}{64}(12m_{1}^2m_{2}^2+3m_{1}^4+5m_{2}^4)
\hspace{1ex} \Biggr )
\labbel{add}
\end{eqnarray}

where

\begin{eqnarray}
t = \frac{ \sqrt{p^2+(m_{1}+m_{2})^2} - \sqrt{p^2+(m_{1}-m_{2})^2}}
{\sqrt{p^2+(m_{1}+m_{2})^2} + \sqrt{p^2+(m_{1}-m_{2})^2}} \ .
\end{eqnarray}

It allows us to reduce the third order equation obtained from \Eq{sys4} 
to a relatively simple second order differential equation

\begin{eqnarray}
\frac{\partial^{2}}{\partial (p^2)^{2} } F_{0,0}^{r} &=& -\frac{2}{p^2}
\frac{\partial}{\partial p^2 } F_{0,0}^{r} - \frac{1}{ (p^2)^{2} }
 R(m_{1}^2,m_{2}^2,-p^2)\left(\frac{C^2(n)}{16}\log(t) 
- \frac{1}{2}S_{l}\right) \nonumber\\
&&- \frac{C^2(n)}{32(p^2)^2} 
\biggl[ (m_{1}^2-m_2^2)\log\left(\frac{m_{1}^2}{m_2^2}\right)
-\log(m_{2}^2)p^2- \log(m_{1}^2)p^2+\frac{7p^2}{2} \biggr] \ .
\labbel{eql1}
\end{eqnarray}

 Its integration is elementary giving dilogarithms 
 as the most 'complicated' functions and the solution after all
 integration constants were fixed reads

\begin{eqnarray}
  &&F_0^{(0)}(m_1^2,m_2^2,0,p^2) = F_{0,0}^{r} = \nonumber\\
 &&C^2(n)\Biggr[ \hspace{2ex} \frac{m_{1}^2m_{2}^2\log^2(t)}{16p^2}
+\frac{1}{64p^2} \biggl(
 -m_{1}^2m_{2}^2\log^2\left(t_1^2\right)
+(m_{1}^4-m_{2}^4)\log\left(t_1^2\right)\biggr)
 \nonumber\\
&+&\frac{p^2}{64} \biggl( \log(m_{2}^2)+\log(m_{1}^2)-\frac{13}{2} \biggr)
-\frac{\log(t)}{32(t-t_{1})} \biggl( \frac{m_{1}^3}{m_{2}}+m_{1}m_{2} \biggr)
 \nonumber \\
&-&\frac{\log(t)}{32(t-t_{2})} 
\biggl( \frac{m_{2}^3}{m_{1}}+m_{1}m_{2} \biggr)
 -\frac{\log(t)}{32}\biggl(R(m_{1}^2,m_{2}^2,-p^2)+2m_{1}^2+2m_{2}^2\biggr)
\nonumber\\
&+&\frac{\log^2(t)-1}{32}(m_{1}^2+m_{2}^2)+\frac{m_{2}^2-m_{1}^2}{16}
\biggl(\Li2(1-\frac{t}{t_{1}})-\Li2(1-\frac{t}{t_{2}}) \biggr)
 \nonumber\\
&+&\frac{2\log(t_{1})(m_{1}^2-m_{2}^2)+m_{1}^2+m_{2}^2}{32}
\log\biggl(\frac{(t_{1}-t)(t_{2}-t)}{p^2}\biggr)\nonumber\\
&+& \frac {\log(m_{1}^2)}{64}(13m_{1}^2+m_{2}^2)
           +\frac {\log(m_{2}^2)}{64}(13m_{2}^2+m_{1}^2)  
           -\frac {\log^2(m_{1}^2)}{128}(5m_{1}^2+m_{2}^2)\nonumber\\
           &-&\frac {\log^2(m_{2}^2)}{128}(5m_{2}^2+m_{1}^2) 
           -\biggl(\frac {\log(m_{1}^2)\log(m_{2}^2)}{64}
           +\frac {5}{32}\biggr)(m_{1}^2+m_{2}^2)
\hspace{2ex} \Biggr] \ \ ,
 \labbel{res}       
\end{eqnarray}

\noindent 
where

\begin{eqnarray}  
t_1=\frac{m_1}{m_2} \ \ \ {\hbox{ \rm and}} \ \ \ t_{2}=\frac{m_2}{m_1} \ .
\nonumber
\end{eqnarray}

The constant \(S_l\) was fixed using the analytically known \cite{CCLR1}
  result for  \(F_0(n,0,0,m_3^2,p^2)\), while two unknown functions of
\(m_1\) and \(m_2\) coming from the integration of the second order 
 equation \Eq{eql1} where fixed from known \cite{CCLR1} results for
 \(F_0(n,m_1^2,m_2^2,m_3^2,0)\) and 
\(\frac{\partial} {\partial p^2}F_0(n,m_1^2,m_2^2,m_3^2,p^2) (p^2=0)\).
 The value of \(p^2\) is to be
 understood, whenever necessary,
 as \(p^2-i\epsilon\), with \(\epsilon\) an infinitesimally small
 positive constant.  

\noindent
 Having 
 \(F_0^{(0)}(m_1^2,m_2^2,0,p^2) \) one can easily get 
 other two independent MI: 
 \(F_1^{(0)}(m_1^2,m_2^2,0,p^2)\) and
 \(F_2^{(0)}(m_1^2,m_2^2,0,p^2)\). We present here only one of them as
 the other one can be found by an exchange of the masses \(m_1\) and \(m_2\).
 It reads

\begin{eqnarray}
&&\kern-30pt F_1^{(0)}(m_1^2,m_2^2,0,p^2) 
 = - \frac{\partial}{\partial{m_{1}^2}}F_{0,0}^{r} =
\nonumber \\
&-&C^{2}(n)\Biggl[\hspace{2ex}
\frac{1}{32}+\frac{R(m_{1}^2,m_{2}^2,-p^2)\log(t)}{16p^{2}}
+\frac{\log(t_{1}^2)}{32p^2}\biggl(m_{1}^2
-\frac{m_{2}^2}{2}(\log(t_{1}^2)+1)\biggr)\nonumber\\
&+&\frac{\log(t)}{32}\biggl(t_{2}^2-2+\frac{\log(t)}{p^2}(p^2+2m_{2}^2)\biggr)
-\frac{t_{2}^2}{64}\biggl(\log(m_{1}^2)+\log(m_{2}^2)\biggr)\nonumber \\
&-&\frac{\log(m_{2}^2)}{128}\biggl(2\log(m_{1}^2)+\log(m_{2}^2)\biggr)
+\log(m_{1}^2)\biggl(\frac{1}{8}-\frac{5\log(m_{1}^2)}{128}\biggr)
 \nonumber\\
&+&\frac{1}{32}(2-t_{2}^2+\log(t_{1}^2))
 \log\biggl(\frac{(t_{1}-t)(t_{2}-t)}{p^2}\biggr)
+\frac{1}{16}\biggl(\Li2(1-\frac{t}{t_{2}})-\Li2(1-\frac{t}{t_{1}})\biggr)
 \Biggr]  \ \ .
 \nonumber\\
\labbel{F1_res}
\end{eqnarray}

 We have tested numerically the results of \Eq{res} and \Eq{F1_res}
 against results
 obtained by newly developed program \cite{CCR3}, which solves
 numerically the system of differential equations \Eq{a5}. In the program
 one cannot put however one of the masses to zero, 
 though very small values of the masses are allowed.
 Putting sufficiently small masses to eliminate its influence on the result
 an agreement of 10 decimal digits was found between the numerical program
 and the analytical results (\Eq{res} and \Eq{F1_res}).

 \section{Conclusions.}

 An extensive use of the differential equation method allowed to find
 closed analytical expressions for the sunrise master integrals 
 with one mass equal to zero, two arbitrary masses and an arbitrary
 external four momentum.  

\vskip 0.4 cm

%%%%%%%%%%%%%%%%%%%%%%%%%%%%%%%%%%%%%
{\bf Acknowledgments.}

One of us (HC) is grateful to M. Caffo and E. Remiddi for
 valuable discussions concerning the sunrise graph.

%%%%%%%%%%%%%%%%%%%%%%%%%%%%%%%%%%%%%%%%%%%%%%%%%%%%%%%%%%%%%%%%%%%%%%%% 
\def\NP{{\sl Nucl. Phys.}} 
\def\PL{{\sl Phys. Lett.}} 
\def\PR{{\sl Phys. Rev.}} 
\def\PRL{{\sl Phys. Rev. Lett.}} 
\def\NC{{\sl Nuovo Cim.}}
\def\APP{{\sl Acta Phys. Pol.}}
\def\ZP{{\sl Z. Phys.}}
\def\MPL{{\sl Mod. Phys. Lett.}} 
\def\EPJ{{\sl Eur. Phys. J.}} 
\def\IJMP{{\sl Int. J. Mod. Phys.}}


\begin{thebibliography}{99}

\bibitem{TkaChet} F.V. Tkachov, \PL {\bf B 100} (1981) 65; 
               K.G. Chetyrkin and F.V. Tkachov, \NP {\bf B 192} (1981) 159. 
\bibitem{BBBS} F.A. Berends, M. B\"ohm, M. Buza and R. Scharf, 
               \ZP {\bf C 63} (1994) 227.
\bibitem{CCLR1} M. Caffo, H. Czy{\.z}, S. Laporta and E. Remiddi,
               \NC {\bf A 111} (1998) 365, hep-th/9805118. 
\bibitem{BDU} F.A. Berends, A.I.Davydychev, N.I. Ussyukina, \PL
              {\bf B 426} (1998) 95, hep-ph/9712209.
\bibitem{CCR1} M. Caffo, H. Czy{\.z} and E. Remiddi, \NP {\bf B 581}
               (2000) 274, hep-ph/9912501. 
\bibitem{DS} A.I. Davydychev and V.A. Smirnov, \NP {\bf B 554} (1999) 391,
             hep-ph/9903328.
\bibitem{GP} S. Groote and A.A. Pivovarov, {\NP} {\bf B 580} (2000) 459,
             hep-ph/0003115.
\bibitem{CCR2} M. Caffo, H. Czy{\.z} and E. Remiddi, \NP {\bf B 611} 
               (2001) 503, hep-ph/0103014. 
\bibitem{FKK} J. Fleisher, M.Yu. Kalmykov and A.V. Kotikov, 
              \PL {\bf B 462} (1999) 169, hep-ph/9905249.
\bibitem{AMR} M. Argeri, P. Mastrolia and E. Remiddi, hep-ph/0202123.
\bibitem{Ince} E.L. Ince, {\sl Ordinary differential equations},
            Dover Publications, New York, 1956.
\bibitem{CCR3} M. Caffo, H. Czy{\.z} and E. Remiddi, hep-ph/0203256. 
\end{thebibliography}
\end{document}